\documentclass[aps,prc,twocolumn,showpacs,superscriptaddress,floatfix,10pt]{revtex4-1}
\usepackage{graphicx}
\usepackage{dcolumn}
\usepackage{bm}
\usepackage{amsfonts}
\begin{document}

\title{Microscopic study of Ca$+$Ca fusion}

\author{R. Keser}
\affiliation{Rize University, Science and Arts Faculty, Department of Physics, 53100, Rize, TURKEY}
\affiliation{Department of Physics and Astronomy, Vanderbilt University, Nashville, Tennessee 37235, USA}
\author{A.S. Umar}
\author{V.E. Oberacker}
\affiliation{Department of Physics and Astronomy, Vanderbilt University, Nashville, Tennessee 37235, USA}

\date{\today}

\begin{abstract}

We investigate the fusion barriers for reactions involving Ca isotopes
$\mathrm{^{40}Ca}+\mathrm{^{40}Ca}$, $\mathrm{^{40}Ca}+\mathrm{^{48}Ca}$, and $\mathrm{^{48}Ca}+\mathrm{^{48}Ca}$
using the microscopic time-dependent Hartree-Fock theory coupled with a density constraint.
In this formalism the
fusion barriers are directly obtained from TDHF dynamics.
We also study the excitation
of the pre-equilibrium GDR for the $\mathrm{^{40}Ca}+\mathrm{^{48}Ca}$ system and the associated $\gamma$-ray emission
spectrum. Fusion cross-sections are calculated using the incoming-wave boundary condition approach. We examine the
dependence of fusion barriers on collision energy as well as on the different parametrizations of the Skyrme interaction.
\end{abstract}
\maketitle

\section{Introduction}
The microscopic study of nuclear many-body problem
and the understanding of the nuclear interactions that
reproduce the observed structure and reaction properties
are the underlying challenges of low energy nuclear physics.
In this context,
detailed investigations of the fusion process
will lead to a better understanding of the interplay among the strong, Coulomb,
and weak interactions as well as the enhanced correlations present in these many-body
systems.

Recently, particular experimental attention has been given to fusion reactions involving
Ca isotopes~\cite{Ste09,Jia10,Mon10,Mon11}.
These new experiments supplemented the older fusion data~\cite{Alj84} and extended it to lower sub-barrier energies.
Comparison of the sub-barrier cross-sections with those calculated using standard coupled-channel calculations
suggested a hindrance of the fusion cross-sections at deep sub-barrier energies~\cite{Ste09,Jia10,Mon10}.
One of the underlying reasons for the failure of standard coupled-channel approach is the use of
frozen densities in the calculation of double-folding potentials, resulting in potentials that behave
in a completely unphysical manner for deep sub-barrier energies.
While the outer part
of the barrier is largely determined by the early entrance channel properties of the
collision, the inner part of the potential barrier is strongly sensitive to dynamical
effects such as particle transfer and neck formation.
This has been remedied in part by extensions of the coupled-channel approach to include a repulsive core~\cite{Esb06}
or the incorporation of neck degrees of freedom~\cite{IH07a,IH07b}.
More recent calculations~\cite{Esb10,Mon11,Mis11} using the coupled-channel approach with a repulsive core have provided
much improved fits to the data.
A detailed microscopic study of the fusion process for Ca based reactions
$\mathrm{^{40}Ca}+\mathrm{^{40}Ca}$, $\mathrm{^{40}Ca}+\mathrm{^{48}Ca}$, and $\mathrm{^{48}Ca}+\mathrm{^{48}Ca}$
could provide further insight into the reaction dynamics as well as a good testing ground for
the theory since these isotopes are commonly used in fitting the parameters of the effective nuclear
interactions, such as the Skyrme force.

During the past several years, we have developed a microscopic approach for
calculating heavy-ion interaction potentials that incorporates
all of the dynamical entrance channel effects included in the
time-dependent Hartree-Fock (TDHF) description of the collision
process~\cite{UO06a}. The method is based on the TDHF evolution of
the nuclear system coupled with density-constrained
Hartree-Fock calculations (DC-TDHF) to obtain the ion-ion interaction potential.
The formalism was applied to study fusion cross-sections for
the systems $^{132}$Sn+$^{64}$Ni~\cite{UO07a}, $^{64}$Ni+$^{64}$Ni~\cite{UO08a},
$^{16}$O+$^{208}$Pb~\cite{UO09b}, $^{132,124}$Sn+$^{96}$Zr~\cite{OU10b},
as well as to the study of
the entrance channel dynamics of hot and cold fusion reactions leading
to superheavy element $Z=112$~\cite{UO10a}, and dynamical excitation energies~\cite{UO09a}. In all cases, we have found
good agreement between the measured fusion cross sections and the DC-TDHF results.
This is rather remarkable given the fact that the only input in DC-TDHF is the
Skyrme effective N-N interaction, and there are no adjustable parameters.

In Section~\ref{sec:Formalism} we outline the main features of our microscopic approach, the DC-TDHF method.
In Section~\ref{sec:Formalism} we also discuss
the calculation of ion-ion separation distance, coordinate-dependent mass, calculation
of fusion cross-sections, and giant dipole resonance (GDR) formalism. In Sec.~\ref{sec:results} we present interesting
aspects of the reaction dynamics and compare our results with
experiment and other calculations. In Sec.~\ref{sec:summary} we summarize
our conclusions.

\section{\label{sec:Formalism}Formalism}

\subsection{DC-TDHF Method}
In the DC-TDHF approach~\cite{UO06a}
the TDHF time-evolution takes place with no restrictions.
At certain times during the evolution the instantaneous density is used to
perform a static Hartree-Fock minimization while holding the neutron and proton densities constrained
to be the corresponding instantaneous TDHF densities~\cite{CR85,US85}. In essence, this provides us with the
TDHF dynamical path in relation to the multi-dimensional static energy surface
of the combined nuclear system. The advantages of this method in comparison to other mean-field
based microscopic methods such as the constrained Hartree-Fock (CHF) method are obvious. First,
there is no need to introduce artificial constraining operators which assume that the collective
motion is confined to the constrained phase space: second, the static adiabatic approximation is
replaced by the dynamical analogue where the most energetically favorable state is obtained
by including sudden rearrangements and the dynamical system does not have to move along the
valley of the potential energy surface. In short we have a self-organizing system which selects
its evolutionary path by itself following the microscopic dynamics.
All of the dynamical features included in TDHF are naturally included in the DC-TDHF calculations.
These effects include neck formation, mass exchange,
internal excitations, deformation effects to all order, as well as the effect of nuclear alignment
for deformed systems.
In the DC-TDHF method the ion-ion interaction potential is given by
\begin{equation}
V(R)=E_{\mathrm{DC}}(R)-E_{\mathrm{A_{1}}}-E_{\mathrm{A_{2}}}\;,
\label{eq:vr}
\end{equation}
where $E_{\mathrm{DC}}$ is the density-constrained energy at the instantaneous
separation $R(t)$, while $E_{\mathrm{A_{1}}}$ and $E_{\mathrm{A_{2}}}$ are the binding energies of
the two nuclei obtained with the same effective interaction.
In writing Eq.~(\ref{eq:vr}) we have introduced the concept of an adiabatic reference state for
a given TDHF state. The difference between these two energies represents the internal energy.
The adiabatic reference state is the one obtained via the density constraint calculation, which
is the Slater determinant with lowest energy for the given density with vanishing current
and approximates the collective potential energy~\cite{CR85}.
We would like to
emphasize again that this procedure does not affect the TDHF time-evolution and
contains no free parameters or normalization.

In addition to the ion-ion potential it is also possible to obtain coordinate
dependent mass parameters. One can compute the ``effective mass'' $M(R)$
using the conservation of energy
\begin{equation}
M(R)=\frac{2[E_{\mathrm{c.m.}}-V(R)]}{\dot{R}^{2}}\;,
\label{eq:mr}
\end{equation}
where the collective velocity $\dot{R}$ is directly obtained from the TDHF evolution and the potential
$V(R)$ from the density constraint calculations.
In calculating fusion cross-sections this coordinate-dependent mass is used to
obtain a transformed ion-ion potential as described below.

\subsection{\label{sec:Excitation} Excitation Energy}
The calculation of the excitation energy is achieved by dividing the TDHF
motion into a collective and intrinsic part. The major assumption in
achieving this goal is to assume that the collective part is primarily determined by
the density $\rho(\mathbf{r},t)$ and the current $\mathbf{j}(\mathbf{r},t)$. Consequently,
the excitation energy can be formally written as~\cite{UO09a}
\begin{equation}
E^{*}(t)=E_{TDHF}-E_{coll}\left(\rho(t),\mathbf{j}(t)\right)\;,
\label{eq:ex}
\end{equation}
where $E_\mathrm{TDHF}$ is the total energy of the dynamical system, which is a conserved quantity,
and $E_\mathrm{coll}$ represents the
collective energy of the system. In the next step we break up the collective energy into two parts
\begin{equation}
E_{coll}\left(t\right)= E_{kin}\left(\rho(t),\mathbf{j}(t)\right) + E_{DC}\left(\rho(t)\right)\;,
\end{equation}
where $E_\mathrm{kin}$ represents the kinetic part and is given by
\begin{equation}
E_{kin}\left(\rho(t),\mathbf{j}(t)\right)=\frac{m}{2}\int\;\textrm{d}^{3}r\;\mathbf{j}^2(t)/\rho(t)\;,
\end{equation}
which is asymptotically equivalent to the kinetic energy of the
relative motion, $\frac{1}{2}\mu\dot{R}^2$, where $\mu$ is the
reduced mass and $R(t)$ is the ion-ion separation distance.
The dynamics of the ion-ion separation
$R(t)$ is provided by an unrestricted TDHF run thus allowing us to deduce
the excitation energy as a function of the distance parameter, $E^{*}(R)$.

\subsection{Calculation of $R$}
In practice, TDHF runs are initialized with energies above the Coulomb barrier at some
large but finite separation. The two ions are boosted with velocities obtained by assuming
that the two nuclei arrive at this initial separation on a Coulomb trajectory.
Initially the nuclei are placed such that the point $x=0$ in the $x-z$ plane is the center of mass.
During the TDHF dynamics the ion-ion separation distance is obtained by constructing a dividing plane between the
two centers and calculating the center of the densities on the left and right halves of this
dividing plane. The coordinate $R$ is the difference between the two centers.
The dividing plane is determined by finding the point at which the tails of the two
densities intersect each other along the $x$-axis. Since the actual mesh used in the
TDHF calculations is relatively coarse we use a cubic-spline interpolation to interpolate
the profile in the x-direction and search for a more precise intersection value.
This procedure has been recently
described in Ref.~\cite{DD-TDHF} in great detail.

The standard procedure for calculating $R$ as described above starts to fail
after a substantial overlap is reached (for $R$ values smaller than the ones
studied in this manuscript).
We have observed that if one defines the
ion-ion separation as $R=R_0\sqrt{|Q_{20}|}$, where $Q_{20}$ is the mass quadrupole moment
for the entire system, calculated by using the collision axis as the symmetry axis,
and $R_0$ is a scale factor determined to give the correct initial
separation distance at the start of the calculations. Calculating $R$ this way yields
almost identical results to the previous procedure until that procedure begins to fail
and continues smoothly after that point. Of course the minimum value of $R$ calculated this way is never zero
but is determined by the quadrupole moment of the composite system.

\subsection{Fusion Cross-Section}
We now outline the calculation of the total fusion cross section using an
arbitrary coordinate-dependent mass $M(R)$. Starting from the classical
Lagrange function
\begin{equation}
L(R,\dot{R})=\frac{1}{2} M(R) {\dot{R}^{2}} - V(R) \;,
\label{eq:Lag}
\end{equation}
we obtain the corresponding Hamilton function
\begin{equation}
H(R,P)=\frac{P^2}{2 M(R)} + V(R) \;,
\label{eq:Ham1}
\end{equation}
where the canonical momentum is given by $P=M(R){\dot{R}}$.
Following the standard quantization procedure for the kinetic energy
in curvilinear coordinates~\cite{Sp59}
\begin{equation}
T = \frac{-\hbar^2}{2} \left[ g^{-{\frac{1}{2}}} \frac{\partial}{\partial q^{\mu}}
    g^{{\frac{1}{2}}} g^{\mu \nu} \frac{\partial}{\partial q^{\nu}}\right] \;,
\label{eq:Tgen}
\end{equation}
where $g_{\mu \nu}(q)$ denotes the metric tensor and $g^{\mu \nu}(q)$ the reciprocal tensor,
one obtains the quantized Hamiltonian
\begin{equation}
H(R,\hat{P})=\frac{1}{2} \left[ M(R)^{-{\frac{1}{2}}} \hat{P} M(R)^{-{\frac{1}{2}}} \hat{P} \right] + V(R) \;.
\label{eq:Ham2}
\end{equation}
with the momentum operator $\hat{P} = -i \hbar d/dR$.
The total fusion cross cross-section
\begin{equation}
\sigma_f = \frac{\pi}{k^2} \sum_{L=0}^{\infty} (2L+1) T_L\;,
\label{eq:sigfus}
\end{equation}
can be obtained by calculating the potential barrier penetrabilities $T_L$
from the Schr\"odinger equation for the relative motion coordinate $R$
using the Hamiltonian~(\ref{eq:Ham2}) with an additional centrifugal potential
\begin{equation}
\left [ H(R,\hat{P}) + \frac{\hbar^2 L(L+1)}{2 M(R) R^2}
 - E_\mathrm{c.m.} \right] \psi_L(R) = 0 \;.
\label{eq:Schroed1}
\end{equation}

Alternatively, instead of solving the Schr\"odinger equation with coordinate dependent
mass parameter $M(R)$ for the heavy-ion potential $V(R)$, we can instead use the constant
reduced mass $\mu$ and transfer the coordinate-dependence of the mass to a scaled
potential $U(\bar{R})$ using the well known coordinate scale transformation~\cite{GRR83}
\begin{equation}
d\bar{R}=\left(\frac{M(R)}{\mu}\right)^{\frac{1}{2}}dR\;.
\label{eq:mrbar}
\end{equation}
Integration of Eq.~(\ref{eq:mrbar}) yields
\begin{equation}
\bar{R}= f(R) \ \ \ \Longleftrightarrow \ \ \ R=f^{-1}(\bar{R})\;.
\label{eq:rbar}
\end{equation}
As a result of this point transformation, both the classical
Hamilton function, Eq.~(\ref{eq:Ham1}), and the corresponding quantum mechanical
Hamiltonian, Eq.~(\ref{eq:Ham2}), now assume the form
\begin{equation}
H(\bar{R},\bar{P})=\frac{\bar{P}^2}{2 \mu} + U(\bar{R}) \;,
\label{eq:Ham3}
\end{equation}
and the scaled heavy-ion potential is given by the expression
\begin{equation}
U(\bar{R}) = V(R) = V(f^{-1}(\bar{R})) \;.
\label{eq:Urbar}
\end{equation}

The fusion barrier penetrabilities $T_L(E_{\mathrm{c.m.}})$
are obtained by numerical integration of the two-body Schr\"odinger equation
\begin{equation}
\left[ \frac{-\hbar^2}{2\mu}\frac{d^2}{d\bar{R}^2}+\frac{L(L+1)\hbar^2}{2\mu \bar{R}^2}+U(\bar{R})-E\right]\psi=0\;,
\label{eq:xfus}
\end{equation}
using the {\it incoming wave boundary condition} (IWBC) method~\cite{Raw64}.
IWBC assumes that once the minimum of the potential is reached fusion will
occur. In practice, the Schr\"odinger equation is integrated from the potential
minimum, $R_\mathrm{min}$, where only an incoming wave is assumed, to a large asymptotic distance,
where it is matched to incoming and outgoing Coulomb wavefunctions. The barrier
penetration factor, $T_L(E_{\mathrm{c.m.}})$ is the ratio of the
incoming flux at $R_\mathrm{min}$ to the incoming Coulomb flux at large distance.
Here, we implement the IWBC method exactly as it is
formulated for the coupled-channel code CCFULL described in Ref.~\cite{HR99}.
This gives us a consistent way for calculating cross-sections at above and below
the barrier energies.

\subsection{GDR Excitation}

Let us now consider a central collision in $x$-direction and introduce
the quantity
\begin{equation}
\label{eq:dt}
D(t)=\frac{NZ}{A} \left [\frac{1}{Z} \sum_{p=1}^Z <x_p(t)> -
                                    \frac{1}{N} \sum_{n=1}^N <x_n(t)> \right ]
\end{equation}
which represents the expectation value of the $x$-component of the dipole operator
$d_x / e$ taken with the time-dependent TDHF Slater determinant $|\Phi(t)>$.
Following the bremsstrahlung approach developed by Baran et al.~\cite{B96,Bar09} we
define the dipole acceleration
\begin{equation}
D''(t) = \frac{d^2 D(t)}{dt^2}
\end{equation}
and introduce its Fourier transform
\begin{equation}
D''(\omega) = \int_{t_{min}}^{t_{max}} D''(t) e^{i \omega t} dt\;.
\end{equation}
Alternatively, for nearly harmonic vibrations one can use the expression $\left|D''(\omega)\right|^2 = \omega^4\left|D(\omega)\right|^2$, with
\begin{equation}
  D(\omega) =
  \int_{t_\mathrm{min}}^{t_\mathrm{max}} D(t) e^{i \omega t}
  \sin^4\left(\pi\frac{t-t_{\mathrm{min}}}{t_{\mathrm{max}}-t_{\mathrm{min}}}\right)dt \;.
\end{equation}
The time filtering $\sin^4$ is used to smooth out peaks coming from finite integration time.
The ``power spectrum'' of the electric dipole radiation is given by~\cite{B96}
\begin{equation}
\label{eq:yield}
\frac{dP}{dE_{\gamma}}=\frac{2 \alpha} {3 \pi E_{\gamma}}
 \left |\frac{1}{c} D''(\omega) \right |^{2} \ ,
\end{equation}
where $\alpha = e^2/(\hbar c) \approx 1/137$ denotes the fine structure constant.
Recently, pre-equilibrium GDR excitation has also been studied in the context of TDHF~\cite{SCh07}.

\section{\label{sec:results} Results}
Calculations were done in 3-D geometry and using the full Skyrme interaction
including all of the time-odd terms in the mean-field Hamiltonian~\cite{UO06}.
The primary Skyrme parametrization used was SLy4~\cite{CB98} but we have also
tested the new UNEDF0~\cite{UNE0} and UNEDF1~\cite{UNE1} parametrizations.
For the
reactions studied here, the lattice spans $48$~fm along the collision axis and $15$~fm in
the other two directions. Derivative operators on
the lattice are represented by the Basis-Spline collocation method. One of the major
advantages of this method is that we
may use a relatively large grid spacing of $1.0$~fm and nevertheless achieve high numerical
accuracy.
The initial separation of the two nuclei is $18$~fm for central collisions. The time-propagation
is carried out using a Taylor series expansion (up to orders $10-12$) of the unitary mean-field propagator,
with a time-step $\Delta t = 0.4$~fm/c.
We have performed density constraint calculations every $10-20$ time steps.
The accuracy of the density constraint calculations is
commensurate with the accuracy of the static calculations.

\subsection{$^{40}$Ca+$^{40}$Ca System}

In this subsection we will present our results for the $^{40}$Ca+$^{40}$Ca system.
Most of our general discussions will be provided here as they would be the same
for other systems. Specific points about individual systems and comparison of
results for the three systems will be taken up in the subsequent sections.
\begin{figure}[!htb]
\includegraphics*[width=8.6cm]{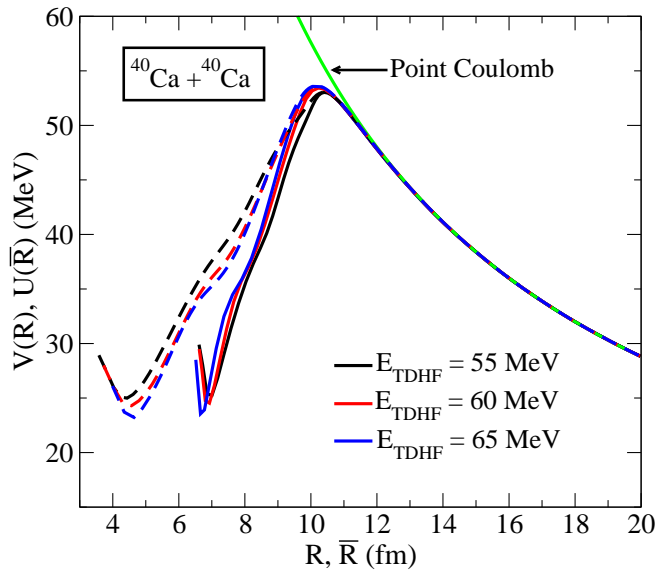}
\caption{\label{fig:fig1} (Color online)
Potential barriers, $V(R)$, for the $^{40}$Ca+$^{40}$Ca system obtained from density constrained
TDHF calculations using Eq.~\protect(\ref{eq:vr}) at three different energies,  $E_\mathrm{TDHF}=55$~MeV
(black solid curve), $E_\mathrm{TDHF}=60$~MeV
(red solid curve), and $E_\mathrm{TDHF}=65$~MeV (blue solid curve). The three dashed curves
correspond to the transformed potential in Eq.~\protect(\ref{eq:Urbar}) using the coordinate dependent
masses. Also shown is the point Coulomb potential.}
\end{figure}

In Fig.~\ref{fig:fig1} we show the microscopic ion-ion potential barriers obtained using Eq.~(\ref{eq:vr})
calculated at three different collision energies, $E_\mathrm{TDHF}=55$~MeV (black solid curve),
$E_\mathrm{TDHF}=60$~MeV (red solid curve), and
$E_\mathrm{TDHF}=65$~MeV (blue solid curve).
We observe a relatively small dependence on the collision energy.
At lower energies the system has more time available for rearrangements to take place
through the formation of a neck, whereas this is less and less the case at higher energies and
the potential barrier approaches the frozen-density limit~\cite{DD-TDHF}.
This produces the observed trend in Fig.~\ref{fig:fig1}, where the lowest barrier peak
corresponds to the lowest energy and the barrier height increases with increasing
collision energy.
The barrier heights are $53.02$, $53.43$, and $53.57$~MeV, respectively.
Similar energy dependence was also observed
in the DD-TDHF calculations of Ref.~\cite{DD-TDHF} and becomes more prevalent for
heavier systems. Similarly, the position of the barrier peak moves toward smaller
$R$ values, albeit very slowly in this case, with increasing energy.
Corresponding $R$ values for the barrier maximum are $10.41$, $10.32$, and $10.23$~fm.
What is also shown on Fig.~\ref{fig:fig1} is the Coulomb potential assuming the two
nuclei to be point particles with $Z=20$. During the approach phase the microscopically
calculated DC-TDHF potential traces the point Coulomb potential, differing by less than
$150$~keV, which provides a test for the numerical accuracy.
\begin{figure}[!htb]
\includegraphics*[width=8.6cm]{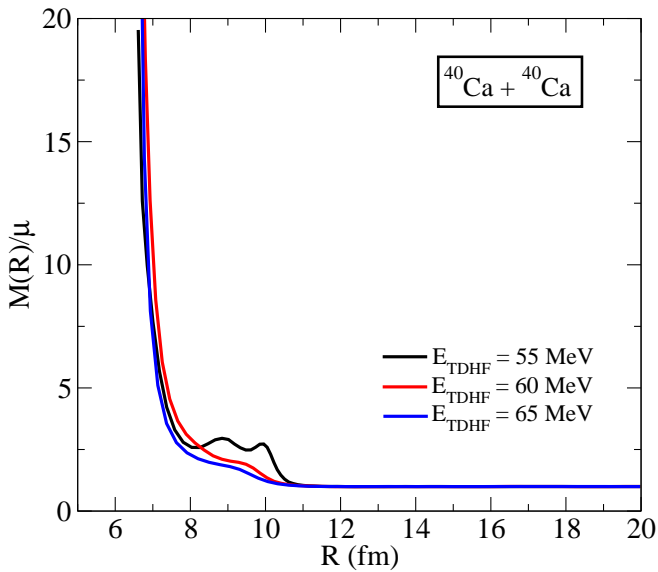}
\caption{\label{fig:fig2} (Color online) Coordinate dependent mass $M(R)$ scaled by the
constant reduced mass $\mu$, obtained from
Eq.~\protect(\ref{eq:mr}), at three different TDHF energies.}
\end{figure}

The energy dependence of potential barriers can also be understood if we examine the coordinate dependent mass of
Eq.~(\ref{eq:mr}) shown in Fig.~\ref{fig:fig2} for three different energies, $E_\mathrm{TDHF}=55$~MeV
(black solid curve), $E_\mathrm{TDHF}=60$~MeV
(red solid curve), and $E_\mathrm{TDHF}=65$~MeV (blue solid curve).
The $R$-dependence of this mass at lower energies is
very similar to the one found in CHF calculations~\cite{GRR83}.
On the other hand, at higher energies the coordinate dependent mass essentially becomes flat,
which is again a sign that most dynamical effects are contained at
lower energies.
The peak at small $R$ values is
due to the fact that the center-of-mass energy is above the barrier and the
denominator of Eq.~(\ref{eq:mr}) becomes small due to the slowdown of the ions.
We have used the coordinate dependent masses shown in Fig.~\ref{fig:fig2} to obtain the
scaled potentials $U(\bar{R})$ of Eq.~(\ref{eq:Urbar}). These potentials are shown
as the dashed curves in Fig.~\ref{fig:fig1}. As we see the coordinate dependent mass
only changes the inner parts of the barriers for all energies. Furthermore, the
effect is largest for the lowest energy collision and diminishes as we increase
the collision energy.
\begin{figure}[!htb]
\includegraphics*[width=8.6cm]{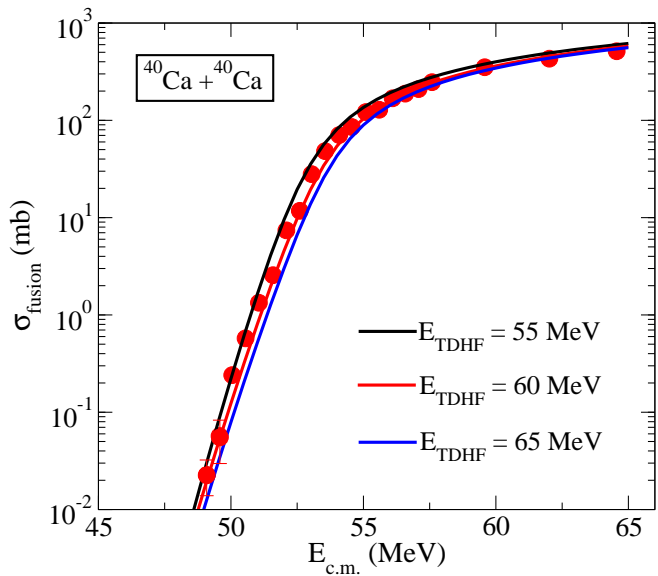}
\caption{\label{fig:fig3} (Color online)
Total fusion cross section as a function of $E_{\mathrm{c.m.}}$.
Three separate theoretical cross section calculations are shown, based on the energy-dependent
DC-TDHF heavy-ion potentials $V(R)$ at energies
$E_{\mathrm{TDHF}}=55$, $60$, $65$~MeV.
The experimental data (filled circles) are taken from Ref.~\protect\cite{Mon11}.}
\end{figure}

We have obtained the fusion cross sections by numerical integration of Eq.~(\ref{eq:xfus}).
The resulting cross-sections are shown in Fig.~\ref{fig:fig3}.
We observe that all of the scaled barriers give a very good description of the
experimental fusion cross-sections. The high energy part of the fusion cross-sections are
primarily determined by the barrier properties in the vicinity of the barrier
peak. On the other hand sub-barrier cross-sections are influenced by what
happens in the inner part of the barrier and here the dynamics and
consequently the coordinate dependent mass
becomes very important. As we observe in Fig.~\ref{fig:fig3} we get
a very good agreement with experiment for the barriers obtained at the lowest
two collision energies, whereas the $E_\mathrm{c.m.}=65$~MeV curve slightly underestimates
the cross-section at lower energies. Although not shown in Fig.~\ref{fig:fig3} the
cross-sections obtained using the the unscaled potentials $V(R)$ and a constant
reduced mass $\mu$ also agree well with the data at higher energies
but either significantly over-estimate or under-estimate the cross-section at
lower energies due to the absence of the coordinate dependent mass.
In summary,  the calculated fusion cross-sections for the $^{40}$Ca+$^{40}$Ca system
reproduce the experimental cross-sections reasonably well, which is a testament
that TDHF with Skyrme force provides a good description for this collision.
\begin{figure}[!htb]
\includegraphics*[width=8.6cm]{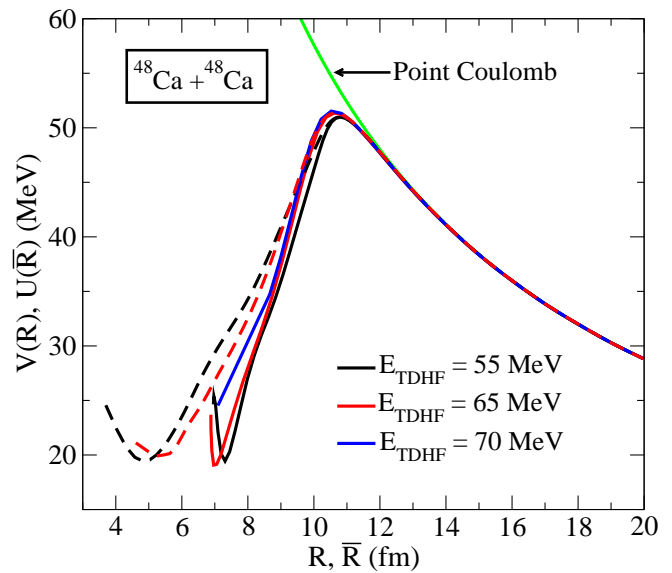}
\caption{\label{fig:fig4} (Color online)
Potential barriers, $V(R)$, for the $^{48}$Ca+$^{48}$Ca system obtained from density constrained
TDHF calculations using Eq.~\protect(\ref{eq:vr}) at three different energies,  $E_\mathrm{TDHF}=55$~MeV
(black solid curve), $E_\mathrm{TDHF}=65$~MeV
(red solid curve), and $E_\mathrm{TDHF}=70$~MeV (blue solid curve). The dashed curves
correspond to the transformed potential in Eq.~\protect(\ref{eq:Urbar}) using the coordinate dependent
masses. Also shown is the point Coulomb potential.}
\end{figure}

\subsection{$^{48}$Ca+$^{48}$Ca System}

In Fig.~\ref{fig:fig4} we show the microscopic ion-ion potential barriers for the
stable neutron-rich $^{48}$Ca+$^{48}$Ca system
calculated at three different collision energies, $E_\mathrm{TDHF}=55$~MeV (black solid curve),
$E_\mathrm{TDHF}=65$~MeV (red solid curve), and
$E_\mathrm{TDHF}=70$~MeV (blue solid curve).
The barrier heights, for increasing collision energy, are $50.98$, $51.37$, and $51.52$~MeV, respectively,
with corresponding $R$ values of $10.80$, $10.63$, and $10.52$~fm.
The comparison of these barriers with those of the $^{40}$Ca+$^{40}$Ca system shows that
the barrier heights are reduced by about $2$~MeV and the location of the barrier maximum
is at a slightly larger $R$ value. This is due to the fact that two two $^{48}$Ca nuclei
are larger than the corresponding $^{40}$Ca nuclei and thus their outer skins come into
contact at a larger $R$ value. After this point the nuclear interaction sets in causing the trajectory
to deviate from the point Coulomb one and producing a peak at a lower energy value.
\begin{figure}[!htb]
\includegraphics*[width=8.6cm]{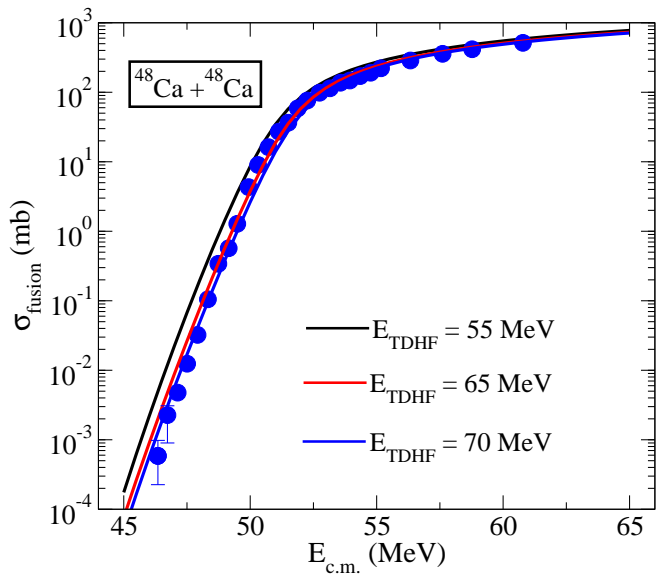}
\caption{\label{fig:fig5} (Color online)
Fusion cross sections for the $^{48}$Ca+$^{48}$Ca system as a function of $E_{\mathrm{c.m.}}$.
Three separate theoretical cross section calculations are shown, based on the energy-dependent
DC-TDHF heavy-ion potentials $V(R)$ at energies
$E_{\mathrm{TDHF}}=55$, $65$, and  $70$~MeV.
The experimental data (filled circles) are taken from Ref.~\protect\cite{Mon11}.}
\end{figure}

Figure~\ref{fig:fig5} shows the fusion cross-sections for the $^{48}$Ca+$^{48}$Ca system
calculated using the barriers of Fig.~\ref{fig:fig4}. While the overall quality of the
agreement with the experimental fusion cross-sections is good, specially for the last two
collision energies, the curve corresponding to the lowest collision energy of $55$~MeV
overestimates the experiment for lower $E_{\mathrm{c.m.}}$ values. This illustrates the
sensitivity of the results to the height of the potential barrier, the difference in this
case between the two barriers being around $0.4$~MeV.
\begin{figure}[!htb]
\includegraphics*[width=8.6cm]{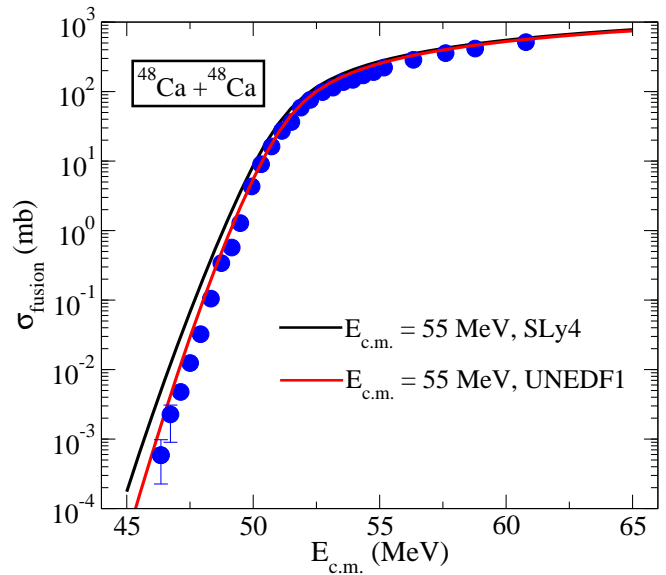}
\caption{\label{fig:fig6} (Color online)
Total fusion cross section as a function of $E_{\mathrm{c.m.}}$ for two different parametrizations of
the Skyrme force, SLy4 and UNEDF1, at the collision energy of $E_{\mathrm{TDHF}}=55$.
The experimental data (filled circles) are taken from Ref.~\protect\cite{Mon11}.}
\end{figure}

While the quality of the DC-TDHF results for the $^{48}$Ca+$^{48}$Ca system is very
good for a parameter free microscopic approach, we have decided to investigate this
further. One of the problems is that the Skyrme fits do badly in reproducing the
single-particle properties~\cite{KD08}, particularly the neutron single-particle states.
For example, with SLy4 parametrization the neutron levels $1d_{5/2}$, $2s_{1/2}$, and $1d_{3/2}$
in $^{48}$Ca nucleus are about $7-3$~MeV
lower in energy than the corresponding experimental values~\cite{UNE0}.
In one of the parametrizations of the nuclear density functional, UNEDF0,
these single-particle energies are raised to values closer to experimental ones.
However, in this case the magic gap at $N=28$ vanishes~\cite{UNE0}.
Recently, a new parametrization, UNEDF1, was introduced with a better incorporation
of large deformations among other improvements~\cite{UNE1}. Another possible advantage of the UNEDF1
parametrization for TDHF calculations is that no center-of-mass correction term was used in the functional,
which is also the case in TDHF.
We have tried this parametrization in our
DC-TDHF calculation for the $^{48}$Ca+$^{48}$Ca system at $E_{\mathrm{c.m.}}=55$~MeV.
In practice, we had to reduce our time-step from $\Delta t = 0.4$~fm/c to $\Delta t = 0.1$~fm/c,
and our density-constraint convergence parameters by a factor of ten or more.
This is probably due to the larger power of the density in the $t_3$ term of the Skyrme interaction.
The peak of the fusion barrier is slightly broader and higher in energy by $187$~keV for UNEDF1.
The use of UNEDF1 results in an improvement of the fusion cross-sections as shown in Fig.~\ref{fig:fig6}.
We have also tried using the parametrization SLy5 but this resulted in no appreciable difference
from the SLy4 case.
\begin{figure}[!htb]
\includegraphics*[width=8.6cm]{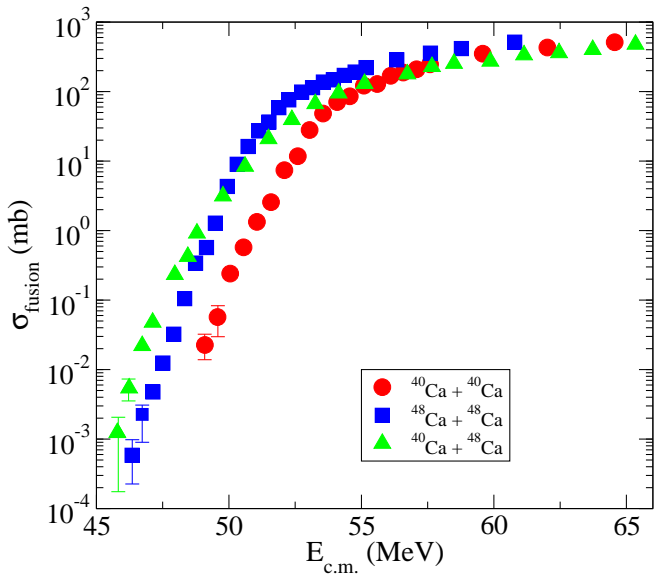}
\caption{\label{fig:fig7} (Color online)
Experimental fusion cross sections as a function of $E_{\mathrm{c.m.}}$ for the three systems studied.
The experimental data are taken from Ref.~\protect\cite{Mon11}.}
\end{figure}

\subsection{$^{40}$Ca+$^{48}$Ca System}

In this section we will examine the fusion of the asymmetric $^{40}$Ca+$^{48}$Ca system.
For highlighting the differences among all three systems we have plotted only the experimental
fusion cross-sections in Fig.~\ref{fig:fig7}. Relative to the $^{40}$Ca+$^{40}$Ca and $^{48}$Ca+$^{48}$Ca
systems the $^{40}$Ca+$^{48}$Ca fusion cross-sections show a different systematic behavior.
At higher bombarding energies the $^{40}$Ca+$^{48}$Ca cross-sections fall below the $^{40}$Ca+$^{40}$Ca data points,
whereas at sub-barrier energies they rise above the $^{48}$Ca+$^{48}$Ca data.
Historically, the theoretical description of the fusion cross-sections for the $^{40}$Ca+$^{48}$Ca system
have been complicated by the couplings to various transfer channels with positive $Q$ values~\cite{LD85,EF89}.
A strong enhancement of fusion cross-sections at lower energies seen in Fig.~\ref{fig:fig7} was attributed to
this effect~\cite{Esb10}.
\begin{figure}[!htb]
\includegraphics*[width=8.6cm]{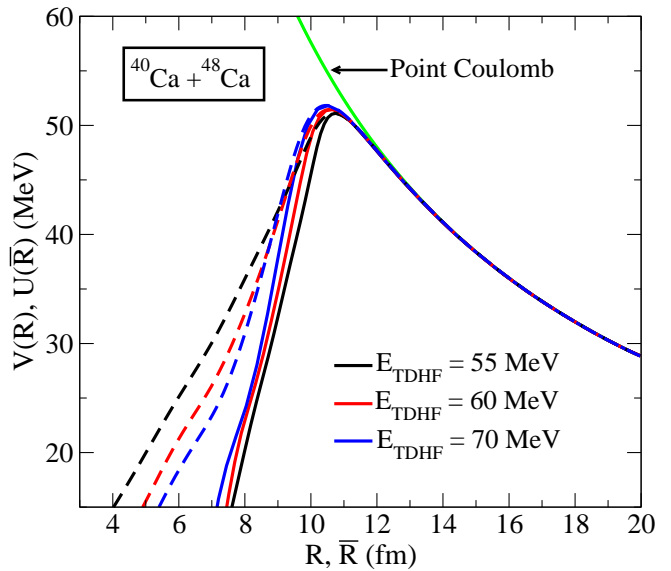}
\caption{\label{fig:fig8} (Color online)
Potential barriers for the $^{40}$Ca+$^{48}$Ca system obtained from density constrained
TDHF calculations at three different energies, $E_\mathrm{TDHF}=55$~MeV
(black solid curve), $E_\mathrm{TDHF}=60$~MeV
(red solid curve), and $E_\mathrm{TDHF}=70$~MeV (blue solid curve). The three dashed curves
correspond to the transformed potential in Eq.~\protect(\ref{eq:Urbar}) using the coordinate dependent
masses. Also shown is the point Coulomb potential.}
\end{figure}

In Fig.~\ref{fig:fig8} we show the microscopic ion-ion potential barriers for the
$^{40}$Ca+$^{48}$Ca system
calculated at three different collision energies, $E_\mathrm{TDHF}=55$~MeV (black solid curve),
$E_\mathrm{TDHF}=60$~MeV (red solid curve), and
$E_\mathrm{TDHF}=70$~MeV (blue solid curve).
The barrier heights, for increasing collision energy, are $51.11$, $51.45$, and $51.81$~MeV, respectively,
with corresponding $R$ values of $10.75$, $10.65$, and $10.51$~fm.
We have used the coordinate dependent masses to obtain the
scaled potentials $U(\bar{R})$ of Eq.~(\ref{eq:Urbar}). These potentials are also shown
as the dashed curves in Fig.~\ref{fig:fig8}. As before the coordinate dependent mass
only changes the inner parts of the barriers for all energies and the
effect diminishes as we increase the collision energy.
\begin{figure}[!htb]
\includegraphics*[width=8.6cm]{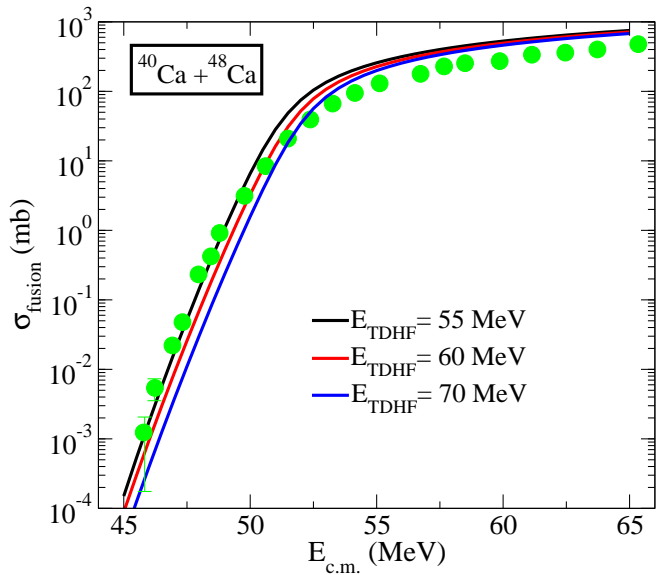}
\caption{\label{fig:fig9} (Color online)
Fusion cross sections for the $^{40}$Ca+$^{48}$Ca system as a function of $E_{\mathrm{c.m.}}$.
Three separate theoretical cross section calculations are shown, based on the energy-dependent
DC-TDHF heavy-ion potentials $V(R)$ at energies
$E_{\mathrm{TDHF}}=55$, $60$, and  $70$~MeV.
The experimental data (filled circles) are taken from Ref.~\protect\cite{Mon11}.}
\end{figure}

Figure~\ref{fig:fig9} shows the fusion cross-sections for the $^{40}$Ca+$^{48}$Ca system
calculated using the barriers of Fig.~\ref{fig:fig8}. The observed trend for sub-barrier
energies is typical for DC-TDHF calculations when the underlying microscopic interaction
gives a good representation of the participating nuclei. Namely, the potential barrier
corresponding to the lowest collision energy gives the best fit to the sub-barrier cross-sections
since this is the one that allows for more rearrangements to take place and grows the inner part
of the barrier. Considering the fact that historically the low-energy sub-barrier cross-sections
of the $^{40}$Ca+$^{48}$Ca system have been the ones not reproduced well by the standard models,
the DC-TDHF results are quite satisfactory, indicating that the dynamical evolution of the
nuclear density in TDHF gives a good overall description of the collision process.
The shift of the cross-section curve with increasing collision energy is typical.
In principle one could perform a DC-TDHF calculation at each energy above the barrier
and use that cross-section for that energy. However, this would make the computations
extremely time consuming and may not provide much more insight.

The trend at higher energies is atypical. The calculated cross-sections
are larger than the experimental ones by about a factor of two.
Such lowering of fusion cross-sections with increasing collision energy
is commonly seen in lighter systems where various inelastic channels,
clustering, and molecular formations are believed to be the contributing
factors~\cite{DDKS}. In the recent coupled-channel approach this is solved
by the addition of a small imaginary potential near the minimum of the
repulsive core~\cite{Esb08}. Such an imaginary part was found not to be
necessary for the $^{40}$Ca+$^{40}$Ca and $^{48}$Ca+$^{48}$Ca systems~\cite{Esb10,Mis11}.
At this time we do not have access to coupled-channel results for the $^{40}$Ca+$^{48}$Ca system.
We have repeated our DC-TDHF calculations using the UNEDF1 interaction which
resulted in a small improvement, reducing the difference with the experimental
values to a factor of about $1.5$. This issue will be discussed further in
the next Section where we examine the excitation properties obtained from TDHF
calculations.

\section{\label{excitation}Excitations}

Excitations are believed to have a significant impact on the outcome of the
fusion reactions. The excitations can range from the entrance channel
quantal excitations of the projectile and target, as in the coupled-channel
approach, to collective excitations of pre-equilibrium system, to compound
nucleus excitations. These can be further influenced by particle transfer,
pre-equilibrium emissions, and evaporation, among others. Theoretically such
effects are commonly introduced by hand into various reaction models.
However, the influence of excitations on nuclear reaction dynamics remains
to be a difficult and an open problem as it combines both nuclear structure
and dynamics under nonequilibrium conditions.
\begin{figure}[!htb]
\includegraphics*[width=8.6cm]{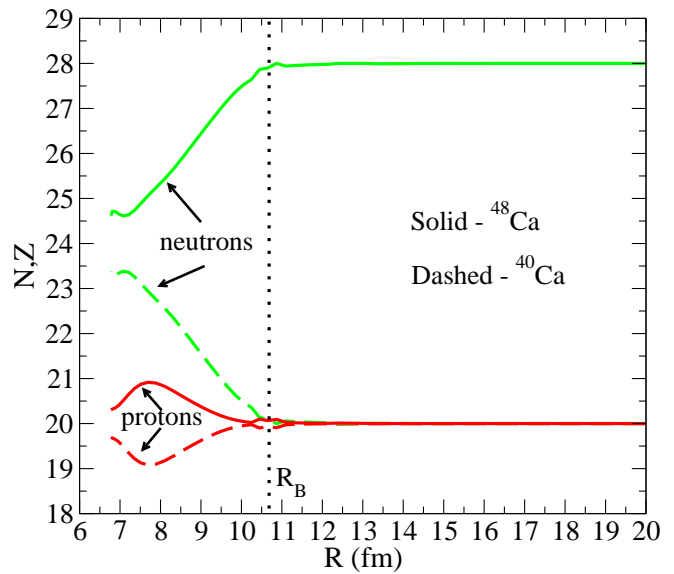}
\caption{\label{fig:fig10} (Color online)
Neutron and proton transfer as a function of ion-ion separation
distance $R$ for the $^{40}$Ca+$^{48}$Ca system. Solid lines denote the particles
originally belonging to the $^{48}$Ca nucleus and dashed lines to the $^{40}$Ca nucleus.
The dotted vertical line shows the location of the potential barrier peak, $R_B$.}
\end{figure}

In Sec.~\ref{sec:Excitation} we have outlined the calculation of the dynamical
excitation energy within the DC-TDHF formalism. This approach was used to
calculate excitation energies for various systems, including the excitation
energy of the heavy systems leading to superheavy formations~\cite{UO10a}. Here,
we have shown that what is also important is the excitation of the system
at the point of capture, which is a deciding factor for forming a composite
system or fusion-fission. This is different than excitation of the compound
nucleus, which determines the survival of the system from quasi-fission.
The dynamical excitation energies calculated from DC-TDHF are relative to
an unequilibrated composite system rather than a true compound nucleus.
\begin{figure}[!htb]
\includegraphics*[width=8.6cm]{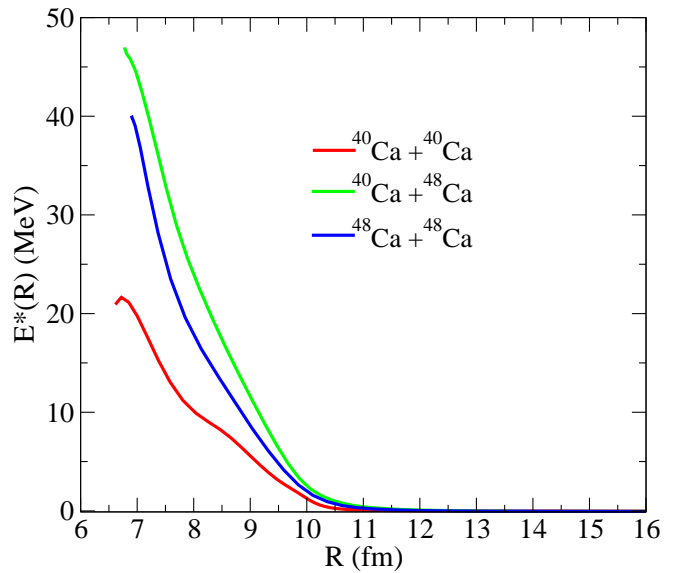}
\caption{\label{fig:fig11} (Color online)
Excitation energy, $E^*(R)$, as a function of the ion-ion separation
distance, $R$, for the three systems studied here.}
\end{figure}

The systems studied in this manuscript present interesting target-projectile
combinations for studying excitations. For example, the non-symmetric
$^{40}$Ca+$^{48}$Ca system will have a pre-equilibrium GDR excitation
in comparison to the other two symmetric systems, as well as particle
transfer. Naturally, some of these effects are included in the fusion
barriers obtained via the DC-TDHF method, as they influence the change
in the TDHF density used in the density constraint calculation.
The quantity $E^{*}(R(t))$ calculated from via TDHF and DC-TDHF represents
the average of dynamical excitations present in the mean-field theory,
with the exception of excitations not determined by the nuclear density
and current (such as the spin currents). Another point to consider is the
pairing interaction. It has been a general consensus that during a
heavy-ion collision the effects of pairing are washed away due to
the high excitations. In this particular case paring effects are minimal
for the initial Ca nuclei studied. However, even if the pairing can
be ignored during the collision process, in many cases it plays an important
role for obtaining good initial HF states as well as obtaining realistic
density constraint solutions when a single composite is formed.
Fortunately, the latter case corresponds to the region of barrier minimum
and not the region around the barrier peak where most fusion cross-sections
are measured. This may in some way explain the success of the DC-TDHF barriers
for fusion.
\begin{figure}[!htb]
\includegraphics*[width=8.5cm]{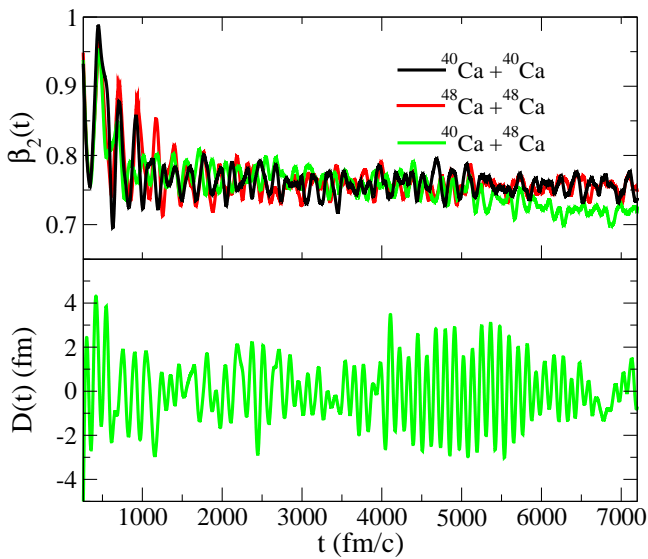}
\caption{\label{fig:fig12} (Color online)
Time evolution of isoscalar deformation parameter $\beta_{2}$ for the head-on collision of
all three systems at a collision energy of $E_\mathrm{TDHF}=55$~MeV (top panel), and the
time evolution of the isovector dipole amplitude $D(t)$ for the $^{40}$Ca+$^{48}$Ca system (bottom panel).}
\end{figure}

In Fig.~\ref{fig:fig10} we plot the average number of neutrons and protons
transferred during the early stages of the TDHF collision of the $^{40}$Ca+$^{48}$Ca system at
$E_{\mathrm{TDHF}}=60$~MeV. The solid lines denote the neutrons and
protons asymptotically belonging to the $^{48}$Ca nucleus and dashed lines
to the $^{40}$Ca nucleus. A number of interesting things can be observed 
from the plot; first is the fact that most of the transfer seems to
start after we pass the potential barrier peak. This indicates that
particle transfer primarily modifies the inner part of the barrier
and not so much the barrier height. The other observation is that
on average about three neutrons are transferred from $^{48}$Ca to
$^{40}$Ca but there is also a small amount of proton transfer in the opposite
direction.
\begin{figure}[!htb]
\includegraphics*[width=8.6cm]{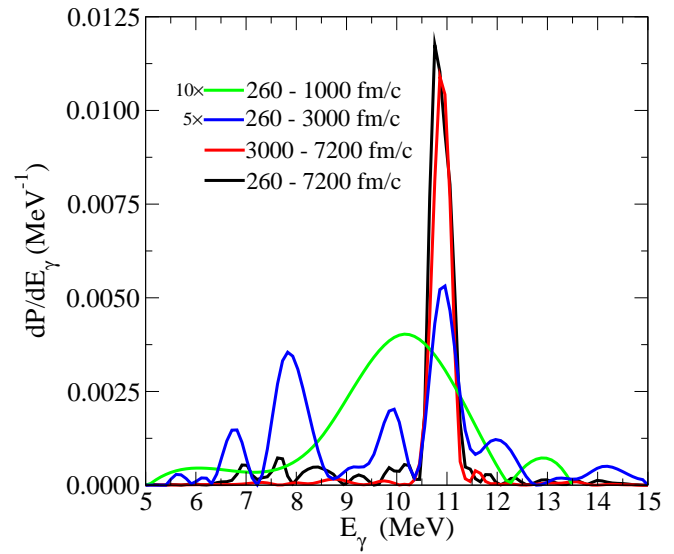}
\caption{\label{fig:fig13} (Color online)
Power spectrum of the isovector dipole amplitude $D(t)$ for the $^{40}$Ca+$^{48}$Ca system is
shown for various time-intervals }
\end{figure}

Figure~\ref{fig:fig11} shows the excitation energy, $E^*(R)$, for the three systems
studies here. The excitation energy was calculated for the same value of $\varepsilon=E_\mathrm{c.m.}/\mu=2.75$~MeV
for all systems, which corresponds to collision energies of $55$, $60$, and $66$~MeV, respectively.
All curves initially behave in a similar manner, at large distances the excitation is zero, as the
nuclei approach the barrier peak the excitations start and monotonically rise for larger overlaps.
The interesting observation is that the excitations for the intermediary $^{40}$Ca+$^{48}$Ca system
start at a slightly earlier time and rise above the other two systems.
This may be largely due to the fact that an asymmetric system has some additional modes
of excitation in comparison to the other two symmetric systems. 
It may be plausible to consider the direct influence of the excitation energy, $E^*(R)$,
on the fusion barriers by making an analogy with the coupled-channel approach and
construct a new potential $V^*(R)=V(R)+E^*(R)$, which has all the excitations added
to the ion-ion potential $V(R)$ that should be calculated at higher energies to 
minimize the nuclear rearrangements (frozen-density limit). 
The resulting potentials somewhat resemble the repulsive-core coupled-channel
potentials of Ref.~\cite{Esb10}.
This approach does
lead to improvements in cases where most of the excitation energy 
is in the form of collective excitations rather than irreversible
stochastic dissipation (true especially for lighter systems).
For the cases studied here we do see a small improvement for the fusion cross-sections
of the $^{40}$Ca+$^{48}$Ca system.
The viability of this approach requires further examination and will be studied in the future.

One of the excitation modes for the $^{40}$Ca+$^{48}$Ca system,
namely the particle transfer, was already discussed above. The others are various isovector
modes such as the pre-equilibrium GDR excitation. In Fig.~\ref{fig:fig12} we show the
time development of the isoscalar deformation parameter $\beta_{2} =\frac{4\pi}{5}\langle r^2Y_{20}\rangle/Ar_\mathrm{rms}$
(top panel) and the isovector dipole amplitude $D(t)$ of Eq.~(\ref{eq:dt}) (bottom panel).
The isovector amplitudes for the two symmetric systems would appear as a zero-line on this plot.
We have left out the initial approach phase of the collision (about $260$~fm/$c$) from these plots.
The time-evolution of the deformation $\beta_{2}$ is very similar for all three systems except
at larger times the $^{40}$Ca+$^{48}$Ca system moves toward smaller deformation in comparison
to the other two systems. This is again most likely due to increased excitations that drive
the system toward a more compact shape.
The evolution of the dipole operator $D(t)$ shows little damping over a long time interval.
The power spectrum associated with this time time-evolution is shown in Fig.~\ref{fig:fig13}.
We have evaluated the spectrum for different time-intervals. The broadest curve corresponds
to a very early stage of the collision, $260-1000$~fm/$c$, but has the lowest amplitude
(multiplied by $10$ in Fig.~\ref{fig:fig13}). In the second interval, $260-3000$~fm/$c$,
we see the development of various peaks and an increase in amplitude. The spectrum for
the entire time interval, $260-7200$~fm/$c$, is dominated by a sharp peak around
$11$~MeV, which is primarily originating from the later stages of the collision which is
evident if one compares it to the spectrum from the interval $3000-7200$~fm/$c$.

\section{\label{sec:summary}Summary}

In this manuscript we have provided a microscopic study of Ca$+$Ca fusion using
the DC-TDHF approach. These reactions have recently been of considerable interest
for the fusion community with a flurry of phenomenological analyses of the fusion
data. Here, we have provided a microscopic alternative to this analysis.
We have shown that microscopically obtained ion-ion potentials do give a 
reasonably good description of the fusion cross-sections.

The fully microscopic TDHF theory has shown itself to be rich in
nuclear phenomena and continues to stimulate our understanding of nuclear dynamics.
The time-dependent mean-field studies seem to show that the dynamic evolution
builds up correlations that are not present in the static theory.
While modern Skyrme forces provide a much better description of static nuclear properties
in comparison to the earlier parametrizations there is a need to obtain even better
parametrizations that incorporate deformation and reaction data into the fit process.

\begin{acknowledgments}
This work has been supported by the U.S. Department of Energy under grant No.
DE-FG02-96ER40963 with Vanderbilt University. RK thankfully acknowledges The Scientific and
Technological Council of Turkey (TUBITAK, BIDEB-2219) for a partial support.
\end{acknowledgments}


\begin{thebibliography}{99}
\bibitem{Ste09} A. M. Stefanini \textit{et al.}, Phys. Lett. B \textbf{679}, 95 (2009).
\bibitem{Jia10} C. L. Jiang \textit{et al.}, Phys. Rev. C \textbf{82}, 041601 (2010).
\bibitem{Mon10} G. Montagnoli \textit{et al.}, Nucl. Phys. A \textbf{834}, 159c (2010).
\bibitem{Mon11} G. Montagnoli and A. M. Stefanini, EPJ Web Conf. \textbf{17}, 05001 (2011).
\bibitem{Alj84} H. A. Aljuwair \textit{et al.}, Phys. Rev. C \textbf{30}, 1223 (1984).
\bibitem{Esb06} \c{S}. Mi\c{s}icu and H. Esbensen, Phys. Rev. Lett. {\bf 96}, 112701 (2006).
\bibitem{IH07a} Takatoshi Ichikawa, Kouichi Hagino, and Akira Iwamoto, Phys. Rev. C {\bf 75}, 057603 (2007).
\bibitem{IH07b} Takatoshi Ichikawa, Kouichi Hagino, and Akira Iwamoto, Phys. Rev. C {\bf 75}, 064612 (2007).
\bibitem{Esb10} H. Esbensen \textit{et al.}, Phys. Rev. C \textbf{82}, 054621 (2010).
\bibitem{Mis11} \c{S}erban Mi\c{s}icu and Florin Carstoiu, Phys. Rev. C \textbf{84}, 051601(R) (2011).
\bibitem{UO06a} A. S. Umar and V. E. Oberacker, Phys. Rev. C \textbf{74}, 021601(R) (2006).
\bibitem{UO07a} A. S. Umar and V. E. Oberacker, Phys. Rev. C {\bf 76}, 014614 (2007).
\bibitem{UO08a} A. S. Umar and V. E. Oberacker, Phys. Rev. C {\bf 77}, 064605 (2008).
\bibitem{UO09b} A. S. Umar and V. E. Oberacker, Eur. Phys. J. A \textbf{39}, 243 (2009).
\bibitem{OU10b} V. E. Oberacker, A. S. Umar, J. A. Maruhn, and P.-G. Reinhard, Phys. Rev. C \textbf{82}, 034603 (2010).
\bibitem{UO10a} A. S. Umar, V. E. Oberacker, J. A. Maruhn, and P.-G. Reinhard, Phys. Rev. C \textbf{81}, 064607 (2010).
\bibitem{UO09a} A. S. Umar, V. E. Oberacker, J. A. Maruhn, and P.-G. Reinhard, Phys. Rev. C \textbf{80}, 041601(R) (2009).
\bibitem{CR85}  R. Y. Cusson, P. -G. Reinhard, M. R. Strayer, J. A. Maruhn, and W. Greiner, Z. Phys. A \textbf{320}, 475 (1985).
\bibitem{US85}  A. S. Umar, M. R. Strayer, R. Y. Cusson, P. -G. Reinhard, and D. A. Bromley, Phys. Rev. C \textbf{32}, 172 (1985).
\bibitem{DD-TDHF} Kouhei Washiyama and Denis Lacroix, Phys. Rev. C {\bf 78}, 024610 (2008).
\bibitem{Sp59}  M.R. Spiegel, {\it Vector Analysis}, Schaum's Outline Series, McGraw-Hill (1959).
\bibitem{GRR83} K. Goeke, F. Gr\"ummer, and P. -G. Reinhard, Ann. Phys. {\bf 150}, 504 (1983).
\bibitem{Raw64} G. H. Rawitscher, Phys. Rev. 135, 605 (1964).
\bibitem{HR99}  K. Hagino, N. Rowley, and A. T. Kruppa, Comp. Phys. Comm. {\bf 123}, 143 (1999).
\bibitem{B96}   V. Baran et al., Nucl. Phys. A \textbf{600}, 111 (1996).
\bibitem{Bar09} V. Baran \textit{et al.}, Phys. Rev. C \textbf{79}, 021603(R) (2009).
\bibitem{SCh07} C. Simenel, Ph. Chomaz, and G. de France, Phys. Rev. C \textbf{76}, 024609 (2007).
\bibitem{UO06}  A. S. Umar and V. E. Oberacker, Phys. Rev. C \textbf{73}, 054607 (2006).
\bibitem{CB98}  E. Chabanat, P. Bonche, P. Haensel, J. Meyer and R. Schaeffer, Nucl. Phys. \textbf{A635},
                          231 (1998); \textbf{A643}, 441(E) (1998).
\bibitem{UNE0}  M. Kortelainen, J. McDonnell, W. Nazarewicz, P.-G. Reinhard, J. Sarich, N. Schunck, M. V. Stoitsov, and S. M. Wild,
                Phys. Rev. C (in press) (http://arxiv.org/abs/1111.4344).
\bibitem{UNE1}  M. Kortelainen, T. Lesinski, J. Mor\'e, W. Nazarewicz, J. Sarich, N. Schunck, M. V. Stoitsov, and S. Wild,
                Phys. Rev. C \textbf{82}, 024313 (2010).
\bibitem{KD08}  M. Kortelainen, J. Dobaczewski, K.Mizuyama, and J. Toivanen, Phys. Rev. C \textbf{77}, 064307 (2008).
\bibitem{LD85}  S. Landowne, C. H. Dasso, R. A. Broglia, and G. Pollarolo, Phys. Rev. C \textbf{31}, 1047 (1985).
\bibitem{EF89}  H. Esbensen, S. H. Fricke, and S. Landowne, Phys. Rev. C \textbf{40}, 2046 (1989).
\bibitem{DDKS}  K. T. R. Davies, K. R. S. Devi, S. E. Koonin, and M. R. Strayer,
                in {\it Treatise on heavy ion Science}, edited by D. A. Bromley,
               (Plenum, New York, 1985), Vol.3, page 3.
\bibitem{Esb08} H. Esbensen, Phys. Rev. C \textbf{77}, 054608 (2008).
\end{thebibliography}
\end{document}